\documentclass[a4paper]{article}

\usepackage[english]{babel}

\usepackage{graphicx}
\usepackage{float}
\usepackage{amsfonts}
\usepackage{multicol}
\usepackage{mathtools}
\usepackage{amssymb}
\def\multiset#1#2{\ensuremath{\left(\kern-.3em\left(\genfrac{}{}{0pt}{}{#1}{#2}\right)\kern-.3em\right)}}
\def\set#1#2{\ensuremath{\left(\genfrac{}{}{0pt}{}{#1}{#2}\right)}}

\usepackage[utf8]{inputenc}
 
\usepackage{amsthm}

\usepackage{geometry}
\newtheorem{innercustomgeneric}{\customgenericname}
\providecommand{\customgenericname}{}
\newcommand{\newcustomtheorem}[2]{%
  \newenvironment{#1}[1]
  {%
   \renewcommand\customgenericname{#2}%
   \renewcommand\theinnercustomgeneric{##1}%
   \innercustomgeneric
  }
  {\endinnercustomgeneric}
}

\newcustomtheorem{customthm}{Theorem}
\newcustomtheorem{customconj}{Conjecture}

\begin{document}
\title{``LOADS of Space" \\ Local Order Agnosticism and Bit Flip Efficient Data Structure Codes\\
\large Senior Thesis 2019}
\author{Matthew Gray}
\maketitle

\begin{multicols}{2}
\begin{abstract}
Algorithms, data structures, coding techniques, and other methods that reduce bit-flips are being sought to best utilize hardware where flipping bits is the dominating cost. Write efficient memories were introduced by Ahlswede and Zhang as a model for storage systems with these kinds of arbitrary read, write, and update costs. The introduction of non-volatile Random Access Memories like phase-change RAM, which have asymmetric read-write costs has re-motivated the field. Our work focuses on potential bit-flip efficiencies to be gained at the data structure layer. We examine Local Order Agnostic Data Structures (LOADS), data structures in which local order does not convey information and in which cells are modified individually. We found that because these data structures have a limited set of valid values and transitions, that bit flipping wins should be possible without the use of additional hardware.
\end{abstract}
\section{Introduction}

In 1989 Ahlswede and Zhang\cite{ahlswede_zhang_1989} introduced the model of write efficient memories. In full generality they a describe memory system with some alphabet of possible characters, some number of slots that can contain any character from the alphabet, and an arbitrary transition cost Matrix that denotes the cost of transitioning any slot between any two characters. However most memory systems have highly symmetric read and write costs for any values. This lack of motivating hardware resulted in minimal follow-up work on the 1989 paper until the emergence of interest in phase-change RAM.

Each bit in a phase-change RAM is stored as one of two crystalline configurations of chalcogenide glass. These two configurations have different resistances and so the configuration can be easily checked by measuring the drop in voltage when running a small current across the bit. However changing the configuration requires running enough current through the bit to melt it and let it recrystallize in the other configuration. So for this memory system the overwhelming cost in terms of power and damage to the hardware comes from the number of bit flips that a computation uses.

The emergence of phase-change RAM has has re-motivated the study of a binary write efficient memories, leading to a slew of recent papers. Many of these papers are concerned with constructing codes using additional coding bits to create multiple codewords representing each possible value. These codes attempt to minimize the Hamming distance between each codeword of each value and the nearest codeword of every other value \cite{jacobvitz2013coset} \cite{Cho2009FlipNWriteAS} \cite{li_jiang_2013}. These codes are now a fairly well plowed field and will be referenced only in passing in this paper.

Instead this paper will be concerned with bit flipping wins that can be found by taking advantage of properties on the data structure layer. We call the main property in question local order agnosticism: where the order of nearby elements can either be inferred from their values or does not matter. This property shows up in two classes of data structures that seem to be polar opposites, data structures that are either sorted or unsorted. In inherently sorted data structures such as hash tables with linear probing, large sets of possible arrangements of values will never be written. In inherently unsorted data structures, such as an unsorted sets, all the rearrangements of the same set of values are all equivalent. In both cases this redundancy can be taken advantage of to allow for write efficient coding ``for free", where ``for free" means with no additional physical memory costs. Encoding and decoding processes may take additional time and computational resources. 

\section{Locality and The General Memory Model}

Previous work on bit flip efficient methods has focused primarily on ways to represent or arrange individual values in such a way that they can be overwritten with other values in a bit flip efficient manner. Our key insight was to take a step back and think about codewords of larger blocks of data. This focus on local neighborhoods allows us to take advantage of local order agnosticism as well as any other structural restrictions on the valid sets of values, or equivalencies between them.

This paper is concerned with the ``local" behavior of data-structures. When we refer to ``locality" or a ``local neighborhood" we are referring to an area of memory containing $k$ slots, where each slot contains values that are $n$ bits long. Within this framework we will have codewords of length $n \times k$, which will decode into a set of up to $k$ values each of which are $n$ bits long. Depending on the memory model we are working under this decoded set may be ordered or unordered.

The least restrictive model over these $n \times k$ bits is the General Memory Model. In this model any possible value over the full $n \times k$ bits is valid and has unique meaning, and transitioning between any two values is allowed. A good example of this might be a Cartesian Point structure in $\mathbb{Z},^k$ with $k$ coordinates being stored as those $k$ integers in order. The important observation here is that the order of the those coordinates conveys which dimension each coordinate varies over. If you rearranged the coordinates you change the point. In the General Memory Model any or all of the coordinates can be changed in a single write. Every set of $k$ coordinates represents a valid Cartesian Point and each Cartesian Point can be written down exactly one way.

\section{Breaking Down the Data Model}

As we examined the linear probing hash table and related data structures, we realized that they had three principal properties.
\begin{itemize}
\item Local Order Agnosticism (LOA) which we also refer to as the Multi-set model: the data structure can be broken up into blocks within which local order does not convey information. There are two main ways a data structure can have this property, either order can be inferred from the data (for instance with a sorted list), or local order does not affect the meaning of the data structure (for instance in an unsorted collection where A,B,C is equivalent to C,B,A).

\item Uniqueness of Elements (UoE) :  the data structure cannot contain duplicate elements within any one block (so A,A,B is an invalid set of data for a block to have).

\item Single Cell Modifications (SCM): during an update to a given block, at most one element will be added or deleted.
\end{itemize}
The first property depends exclusively on the data structure. The second property can be insured by the data structure, for instance hash tables that store key-value pairs will have no duplicates because each key can only appear once, or can be insured by guarantees that the inputs will be unique. 

The third property typically requires assumptions about the memory hierarchy that this program is operating within. If the program is writing directly to non-volatile RAM then this would be a very reasonable assumption, however pretty much all modern computers have some cache hierarchy. In the presence of a cache hierarchy, our work only accurately models programs / data structures with bad cache locality that write to the same block only once before getting flushed from cache to RAM.

Hash tables with linear probing provide a good example of a data structure that will have all three properties even with a cache hierarchy. Losing information about local ordering will not stop us from successfully probing. Because we can't have a key stored with multiple values each key-value pair will be unique. And because each block will be written to with roughly uniform probability, assuming we have a good hash function, with high probability each block will be flushed from the cache with only one modification.

These three properties can be combined to find eight possible memory models from the ``General Memory Model" with none of these properties to the full LOADS memory model. These memory model seem to break down into roughly three groups. The Multi-set, Set, and Ordered Unique models behave optimally under a compression and a classic WEM codes model. SCM and the General Memory Model appear to have no wins that we can take advantage of. Where things appear to get most interesting is in the combination of SCM with uniqueness and / or local order agnosticism. 

\begin{figure}[H]
\centering
\includegraphics[scale=0.4]{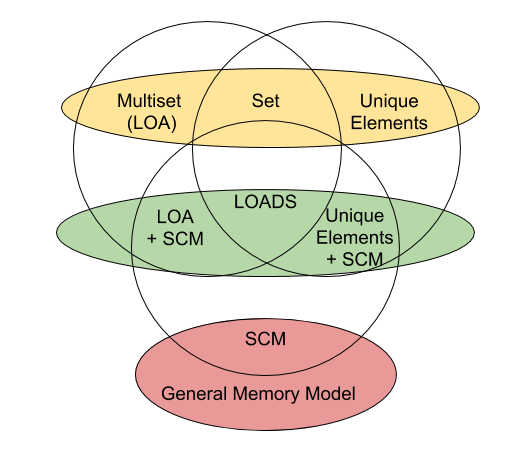}
\caption{The 8 memory models}\label{models}
\end{figure}

Before we get to the interactions of the properties, we will dig a little bit deeper into the  precise ways Local Order Agnosticism and Uniqueness of Elements decrease the number of possible values our data structure needs to be able to represent. 

\section{Properties that Limit the Symbol Space}

In the General Memory Model every bit-string of length $nk$ represents a unique and valid possible value for our block. If we assume our data structure is constrained by uniqueness of elements then any bit-string in which two slots contain the same value is invalid.The number of possible values is given by the below equation.

If our data structure is local order agnostic than any two bit-strings with the same set of slot values are equivalent. The number of unique values is the number of multisets of up to $k$ bit-strings of length $n$, this number is given by given by the below equation. The double parentheses in this equation is the notation for multi-set choose.

 $$ \multiset{2^n}{k}  =  \set{2^n+k-1}{k} = \frac{(2^n+k-1)!}{(k)!(2^n-1)!}$$

If our data structure is both local order agnostic and has unique elements, then the valid / unique elements are all the sets of up to $k$ bit-strings of length $n$, this number is given by the below equation.

 $$ \set{2^n}{k}  =  \frac{(2^n)!}{(k)!(2^n-k)!}$$

For these three models we are dealing with a fairly simple situation in which we have a certain number of valid symbols and $2^{n \times k}$ possible bit-strings that we can use to represent them. Without the addition of single cell modification all transitions between the symbol set are valid. The only difference between this and the generally studied area of write efficient memory is that our number of valid symbols might not be a power of two. We would be surprised if this has a huge effect and one can simply add a collection of "NULL" values to pad out the symbol set to a power of two. Write efficient memory is a fairly well studied field with optimality results proven for polar codes \cite	{li_jiang_2013}. The one other obstacle is compressing the set of symbols down into an efficient representation, however work has been done for both sets and multisets \cite{kovacevic_tan_2018} \cite{steinruecken_2014}. While during this project we decided to not pursue an attempt to combine these two techniques, we believe that this is one of the lowest hanging fruit for future researchers. During this project we saw the juicer but higher hanging fruit of SCM and doggedly failed to pick it. Separately, in the case of the Unique Elements Model, $k$ has to be very high before multiple codewords of every value are possible and so we chose not to concern ourselves with additional study of that memory model.

\section{Analysis of SCM}

Under the General Memory Model any transition between block values is valid. Under the Single Cell Modification Model, only one block may be changed during each write. There are two main variations on this model: first the write/delete model where we can either clear a slot, or if a slot is empty, we can write that slot; and second the overwrite model where any one slot (empty or filled) can be set to any value including empty. In both models only a single slot can be changed in any one write.

We first examine the SCM overwrite model. In this model any of the $2^{n \times k}$ bit-strings is valid and unique, but for any given setting there are only $k \times (2^n-1)$ possible transitions. Each valid transition involves choosing one of the $k$ slots and changing its value to one of the $2^n -1$ values it currently does not have (this includes $n$ 0's which is how we typically encode that the slot is empty in write/delete models). Under the General Memory Model each bit-string has $2^{n \times k}$ possible valid transitions. For building write efficient codes we need to have a much smaller set of codewords nearby any given point.

\subsection{The Hypercube Representation of Bit-strings}

When we say "at a given point" we are implicitly referring to the hypercube model of the world of bit-strings which provides an incredibly helpful visualization and model for understanding many ideas in Coding Theory. To use this model we think of the universe of length $n$ bit-strings as existing at the corners of an $n$-dimensional hypercube.

\begin{figure}[H]
\centering
\includegraphics[scale=0.28]{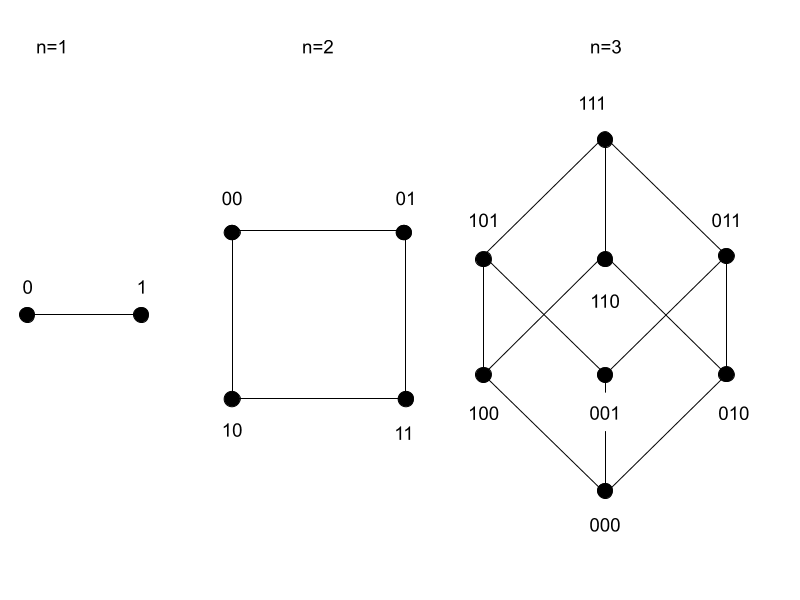}
\caption{The first three dimensional hypercubes.}\label{hypercube}
\end{figure}

Under this model it's clear that the goal of any Write Efficient Code is to have a representation of every state that can be transitioned to at a small Hamming distance away on the hypercube. Hamming distance being the number of bit-flips necessary to move from one bit-string to another. This contrasts with Error Correcting Codes in which the goal is to have no valid representations of any state that are a small Hamming distance away.

\subsection{Information Theoretic Limits}

On the Hypercube the number of ``neighbors" with maximum distance $d$ that each point has is equal to: 
$${\displaystyle \sum_{i=0}^{d}{\set{n \times k}{d}}}$$ 
Under the General Memory Model, we have to be able to transition to any of the $2^{n \times k} - 1$ possible bit-strings. This requires $d$ to equal $n \times k$ meaning that there are transitions we may be forced to take with cost as high as $n \times k$. Under the General Memory Model we may be asked to flip every single bit in our bit-string. Under overwrite SCM though the number of valid transitions is only $k \times (2^n -1)$ which is much smaller than $2^{n \times k} -1$. For example, if $k$ = 4 and $n$ = 4, then we need to be able to transition to $4 \times (2^4-1) = 4 \times (15) = 60$ possible states. Under a trivial encoding we already can only be asked to flip the 4 bits of any one slot, but if we allow ourselves to modify any of our 16 bits there are 136 bit-strings with distance at most 2 which is more than enough to (in theory) have a codeword of every one of those 60 states within distance at most 2. So we have an opportunity to halve our maximum cost. This behavior gives us increased opportunities as the number of slots $k$ increases until $k > 2^n$ at which point every possible transition may be placed at Hamming distance 1. If we also add in Uniqueness of Elements and Local Order Agnosticism we could even have a simple indicator vector with the $i^th$ bit representing if the value $i$ is part of our set. 

\section{Conjectures 1 and 2}

SCM provides some very promising opportunities for bit flip efficiency. However the hypercube is a nasty beast. Under pure SCM (without Local Order Agnosticism, or Uniqueness of Elements) we conjecture that these opportunities cannot be realized. This is because under pure SCM, every value has exactly one codeword which must be placed somewhere in the hypercube. So creating a code is equivalent to doing a series of element swaps. Every swap can decrease the hamming distance for some transitions, but usually at the cost of increasing the hamming distance for many others. From experimenting with very small $n$ \and $k$ we were unable to find an example of a coding mechanism that decreased maximum or average bit-flips.

\begin{customconj}{1}\label{one}
In the SCM model there do not exists codes such that the average or maximum cost of valid transitions is less than that of the trivial encoding. 
\end{customconj}

However we do believe that with the flexibility of having multiple codewords of each value should allow us to realize some of the promise of SCM. We believe that wins should be possible by using built in redundancy. This redundancy can come from guarantees on the data, or be uncovered through compressing the elements left after taking into account Local Order Agnosticism, Uniqueness of Elements, or both.

\begin{customconj}{2}\label{two}
Given an SCM model with sufficient redundancy, there exist codes such that the average or maximum cost of valid transitions is less than the costs using normal encoding methods on each slot individually.
\end{customconj}

\section{Difficulties in proving the conjectures}

Because a pure SCM code is a mapping from bit-strings of length $nk \rightarrow$ bit-strings of length $nk$, it can be thought of as a reordering of the $2^{n \times k}$ bit-strings. There are $(2^{n \times k})!$ such reorderings. This is an absolutely enormous space to be working in even for very very small $n$ and $k$. This makes any sort of brute force proof of Conjecture 1 very difficult even for case studies of small $n$ and $k$. And if Conjecture 1 turned out to be wrong, finding the useful codes within that huge space is a monstrously difficult task without some principled method.

The first proof techniques that we attempted for Conjecture 1 was to argue that each swap made during the creation of an ordering representation of the code would have cost higher than gain. However while we conjecture that the gain at each step can never outweigh the costs imposed by all previous swaps, if some very poor swaps were made previously a single swap could have very positive effects. For instance if we had swapped all elements with their not ($10001$ would be swapped with $01110$) except for the all 0's and all 1's string then swapping those two would get them much closer to their sets of valid neighbors (the one's where all slots but one are 0's or 1's respectively). We failed to come up with a promising technique to prove that the gain at each step can never outweigh the previous costs.

The second proof technique we attempted was inspired by coding theory proofs for optimality of codes such as Huffman codes. This technique works by a clever use of induction and contradiction. The proof goes something like this. Assume code $H_n$ is optimal for any alphabet of size $n$. We then assume that code $H_{n+1}$ is not optimal for an alphabet of size $n+1$. This means there is a $H'_{n+1}$ that is better than $H_{n+1}$, we then use $H'_{n+1}$ to construct a code $H'_n$ that is better than $H_n$. This is a contradiction because by the inductive hypothesis $H_n$ was optimal.

This proof technique transfers over pretty well. We hoped to show that there did not exist a code $C_{n,k}$ with who's total cost function 
$${\displaystyle \sum_{e_1 \in C}\sum_{e_2 \in C}{| e1 \oplus e2 |}} $$
is less than the trivial representation cost of $(2^n)(n \times k/2)$. Because when $k = 1$ we are in the General Memory Model of length $n$, every transition is valid and so there is no way to reduce the total transition cost. So all we need to do is induct over $k$. We'd hoped that the existence of improved codes with $k + 1$ slots would allow for the construction of improved codes with $k$ slots. In one case, where a collection of $k$ bits are affected only by the value in a single cell, these ``Isolated" bits could be removed creating an improved code for $k$ slots. But outside of that special case we struggled to find a way to construct $k$ slot codes from $k+1$ slot codes.

We still believe that Conjecture 1 is true and we suspect some of the tools in the original WEM paper \cite{ahlswede_zhang_1989} might be used to prove this result. However that paper's notation makes many of its main results very unclear and the techniques used to find those results even more so. Further research on this question is recommended.

It seemed like the simplest approach to proving Conjecture 2 would be to find a code with better performance. We failed to do this. The other approach that was considered was a proof by the probabilistic method popularized by Erd\"os\cite{alon2004probabilistic}. For this we would show that a random code would have performance $x$ with probability $p_x$. Then hopefully show that the probability of having non trivial performance was non zero. However we were unable to find a way of calculating the performance of a random code so this approach stalled out. Further research into Conjecture 2 is strongly recommended.

\section{Conclusion}

After this section we include additional appendices on short paths that our research took us down; some attempts at formalizing our two Conjectures; calculations, formulas, and remarks related to the behavior of our various memory models; and advice for those attempting to follow this work. However the core of our results and the narrative of our intuition ends here. 

The vast majority of work in Bit-Flip Efficient Methods has focused on either using redundancy to pack codewords of every symbol closer to each other on the hypercube. The second large body of work concerns using patterns in the data itself to reduce bit-flips. We decided to eschew both these to examine the restrictions on a data structure's locally valid values and transitions. The first upside of this approach turned out to be that we could extract redundancy directly out of a data structure's rules without having to devote extra physical resources to memory. The second upside was to discover the potentially huge wins of the Single Cell Modification model which massively restricts the set of legal transitions over a larger local area. This is an incredibly promising area with the possibility for large wins for bit-flip efficiency that require fairly low computational overhead and no additional hardware costs. However we spent most of our energy focused on the biggest and most interesting potential win: taking advantage of SCM, and failed to prove our two main conjectures about the achievability of those wins, and failed to find codes which took advantage of those wins.

For future work, the lowest hanging fruit is to take a deeper look at the set encoding mechanisms and to see if the redundancy left from set compression could be used for classical error correction and bit-flip efficient codes. This would give significant free wins for data structures including the linear probing hash table. The more difficult but more rewarding research path would be to try to verify Conjectures 1 \and 2. The potential wins here are much larger, but the path forward likely involves the use of serious tools from information theory and coding theory.

\section{Acknowledgements}

I cannot thank my collaborator Justin Raizes enough. He has been an incredible friend for the last four years, and his ideas, inspiration, and excitement made this work possible. I would like to thank my sister Nelle Gray for her help with multi-sets. It's wonderful having family I can reach out to when I find myself in combinatorial hot water. And I would like to thank Andrew Stolman and Daniel Bittman for providing skeptical but encouraging listening ears.

I'd like to thank my advisors, Peter Alvaro, Seshadhri Comandur, and Ethan Miller. All three have served as fonts of inspiration and advice.

And lastly I'd like to thank Darrell Long who first exposed me to NVRAM and bit-flip efficiency. He opened my eyes to the fact that professors are among the most interesting people in the University, and inspired me to go talk to them more.

\appendix
\section{Details on the Data Models}

This section contains the formulas for the number of valid states and transitions under each of the Memory Models.

The properties that effect number of valid values are Local Order Agnosticism and Uniqueness of Elements. We will consider the 4 combinations of these properties. If Single Cell Modifications is added this does not change the number of valid states, only the number of valid transitions. For each of the four underlying memory models we list the number of valid states, as well as the number of valid transitions in the overwrite and write/delete single cell transition models. In this table $s$ represents the number of cells that are set, and $v$ is the number of values that have been written which may different from $s$ because for models without Uniqueness of Elements multiple cells may be set to the same value.

\begin{center}
\begin{tabular}{ |c|c| } 
 \hline
Model & States\\
 \hline
GMM & $2^{(n \times k)}$ \\ 
 \hline
UOE & ${\displaystyle \sum_{i=0}^{k}{\prod_{j=0}^{i}{2^n-j}}}$\\
 \hline
LOA & $\sum_{i=0}^{k}{\multiset{2^n}{k}} = \sum_{i=0}^{k}{\frac{(n+i-1)!}{i!(n-1)!}}$\\ 
 \hline
Set & $\sum_{i=0}^{k}{\set{2^n}{k}} = \sum_{i=0}^{k}{\frac{(n)!}{i!(n-i)!}} $ \\ 
\hline
\end{tabular}
Number of valid states
\end{center}

\begin{center}
\begin{tabular}{ |c|c| } 
 \hline
Model & Overwrite \\
 \hline
GMM & $2^{n \times k}$ \\
 \hline
SCM   & $k \times (2^n-1)$ \\ 
 \hline
SCM + UOE & $k \times (2^n-s)$ \\ 
 \hline
SCM + LOA & $(v+1) \times (2^n-1) < k \times 2^n$ \\ 
 \hline
LOADS   & $(v+1) \times (2^n-1) < k \times 2^n$ \\ 
 \hline
\end{tabular}
Number of valid overwrite transitions
\end{center}

\begin{center}
\begin{tabular}{ |c|c| } 
 \hline
Model & Write/Delete \\
 \hline
GMM & Not defined \\
 \hline
SCM &  $s+(k-s) \times (2^n-1)$\\ 
 \hline
SCM + UOE & $(2^n-1)$\\ 
 \hline
SCM + LOA & $v+(2^n-1)$ \\ 
 \hline
LOADS & $2^n$ \\ 
 \hline
\end{tabular}
Number of valid write/delete transitions
\end{center}

Below are derivations for a few representative entries. The rest follow similarly. We encourage the reader to attempt a few of the derivations if they so wish. 

The number of valid states for Uniqueness of Elements is the number of ordered subsets of the $2^n$ with cardinality at most k. To select one of these first choose $i$ the number of elements you will have in your subset, then choose one of the $2^n$ elements to be your first element, then choose one of the $2^n-1$ elements to be your second. Continue pulling one of $2^n-j$ elements until $j=i$.

The number of valid transitions for LOADS under write/delete cell modification is exactly equal to the number of values a single cell can hold. To select a transition choose a value, either the value will already be in the data structure in which case you delete it. Otherwise you are adding it. It is worth noting that this contains a special case in which the data structure is full with all $k$ cells set. In that case the only valid transitions are deleting one of the $k$ values stored.

The number of valid transitions under SCM, Local Order Agnosticism, and overwrite transitions can be derived as follows. To select an element of the set of valid transitions first choose either one of the $v$ values to overwrite or choose to add a new value. Then choose a value to write to either the empty cell or overwrite an instance of the chosen value with. In the case of an overwrite you cannot write the value already there, and in the case of a fresh write you can't write the all 0's value because that indicates NULL or empty.

The General Memory Model cannot have the write delete property as that property really only makes sense in the context of being able to only write or delete a single cell. For the General Memory Model to have this property we would have to assume we could only clear or write entire blocks, in that case we would either have exactly one possible transition (if there is a value that we must delete), or all $2^{n \times k}$ (if the block is currently empty).

\section{Attempt at Semi-Linear Codes}

One route that seemed promising early on for the full LOADS model was ``Semi-Linear" Codes. Linear codes are a family of codes in which every linear combination of codewords is a codeword in turn. Our ``Semi-Linear" Codes worked slightly differently. For this system any sum of less than $k$ basis codewords would be a codeword. These basis codewords each represent a single slot value, and the combination of up to $k$ such codewords represents the set of the encoded elements. The primary benefits of employing this approach are that encoding and decoding are computationally feasible, and that making arguments about write distances follow very easily from the hamming weights of the values of each single value codeword. 

For a semi-linear code we would have a codeword of length $n \times k$ for each of the $2^n$ possible values. To construct the codeword for some set of slot values we simply xor the values' codewords together. This encoding system has a $2^n \times nk$ encoding Matrix $E$ with the columns $c_0$, $c_1$, $c_2$ that are the basis value's codewords, each of which are $n \times k$ bits long. To encode a set we create an indicator vector $s$ which has $s_i=1$ if value $i$ is in the set. Multiplying this vector by the Matrix will result in the set's codeword.

\[
\begin{bmatrix}
 s_0\\
 \vdots \\
 s_{2^n -1} \\
\end{bmatrix}
\begin{bmatrix}
 c_{0,0} & c_{0,1} &\hdots & c_{0, 2^n -1}\\
 \vdots & \vdots & \ddots & \vdots \\
 c_{nk,1} & c_{nk,1} & \hdots & c_{nk,2^n -1}
\end{bmatrix}
\]
\[
=
\begin{bmatrix}
 \ e_0 & \hdots & e_{nk}\\
\end{bmatrix}
\]

The great thing about this method is that doing the bit flip analysis becomes incredibly simple. The cost of writing or deleting a value is the hamming weight of the value's codeword. Doing an overwrite of one value with another has cost equal to the hamming weight of the xor of those two values' codewords.

We want to guarantee that no two sets $s_1$ and $s_2$ will be encoded such that $e_1 = e_2$. To guarantee this, it is sufficient to guarantee that there is no set of fewer than $2 \times k+1$ column vectors in the encoding matrix which xor to zero. This is equivalent to requiring that any set of $2 \times k+1$ basis vectors are linearly independent. We believe this should be possible but decided to not investigate too much father because we realized that this method cannot be made efficient.

The obvious problem with this method is that storing a matrix with $2^n$ columns is very space inefficient. We hoped that with the addition of a nonlinear function we could implicitly store the encoding Matrix in a much smaller space. This would create a smaller $kn \times m<2^n-1$ matrix $M$. Below we provide a proof that no such function and matrix pair can exist even to just be able to map individual elements to their codewords.

\newtheorem{thm1}{Theorem}
\begin{thm1}
There does not exist any $kn \times m$ matrix $M$ and non linear function $t$ from the natural numbers from $0$ to $2^n -1$ onto vectors in $\mathbb{R}^m$ s.t. $Mt(x) = E\bar{x}$ where $x$ is the natural number from $0$ to $2^n -1$ and $\bar{x}$ is the indicator vector of $x$.
\end{thm1}

$$Mn(x) = E\bar{x} $$

\[
\begin{bmatrix}
M_{0,0} & M_{0,1} &\hdots & M_{0,m}\\
 \vdots & \vdots & \ddots & \vdots \\
 M_{nk,0} & M_{nk,1} & \hdots & M_{nk,m}
\end{bmatrix}
\begin{bmatrix}
 t(x)_0\\
 \vdots \\
  t(x)_{m} \\
\end{bmatrix}
\]
\[
=
\begin{bmatrix}
 c_{0,0} & c_{0,1} &\hdots & c_{0,2^n -1}\\
 \vdots & \vdots & \ddots & \vdots \\
 c_{nk,0} & c_{nk,1} & \hdots & c_{nk,2^n -1}
\end{bmatrix}
\begin{bmatrix}
0 \\
0 \\
\vdots\\
\bar{x}_x = 1 \\
\vdots\\
0\\
0\\
\end{bmatrix}
\]
 
\begin{proof}
Consider the smallest such M (i.e. the one where m is minimized).
Let $F$ be the matrix such that column $x$ is equal to the length m vector $n(x)$.
Now $E$=$MF$.
Because $M$ is the smallest such matrix, $F$ must be full rank. This means that $F$ has a right inverse $F^{-1}_{R}$.
Now if we multiply into the above equation we get $EF^{-1}_{R}$ = $M$

If $M = EF^{-1}_{R}$ then $Mn(x) = EF^{-1}_{R}n(x)$. $F^{-1}_{R}$ is a linear map from an $m$ dimensional space so the subspace it maps too must have dimension at most $m$. However $F^{-1}_{R}n(x)$ must be equal to the the indicator vector that is $0$ at all indexes and $1$ at index $x$. That means that it is mapping to a space that has dimension $2^n-1$. This is a contradiction and therefor $M$ and $t$ cannot exist.
\end{proof}

This impossibility result implies that any efficiently encodable LOADS code must be non-linear w.r.t. the slot values. This is a sobering result as it implies that any coding results in this area may require significantly more intricate constructions.

\section{Attempts at Formalizing Conjectures}

The original WEM paper \cite{ahlswede_zhang_1989} provides powerful tools and language to describe bit-flip efficient methods. And our conjectures can be formalized in terms of this language. Their language is concerned with "rates" which are the ratio of the number of bits needed to give every valid symbol a unique bit-string, and the number of bits used in the codewords for that symbol-set. For $M$ valid symbols and $n$ bits, the rate would be $\frac{log(M)}{n}$. The paper is concerned with proving results on the minimum rate necessary to ensure a maximum or average transition cost.

In their language, the conjectures can be stated as follows.

\begin{customconj}{1}\label{one}
For SCM the minimum distance that can be achieved with rate $1$ is $\frac{1}{k}$.
\end{customconj}

\begin{customconj}{2}\label{two}
Given any non-SCM model besides GMM with symbol set M. Then there exists $n$,$k$ s.t. the minimum distance achieving rate $\frac{log(|M|)}{nk}$ for that model is greater than the minimum distance achieving rate $\frac{log(|M|)}{nk}$ for that model + SCM.
\end{customconj}

We hope that follow up work will be able to extract tools from Ahlswede and Zhang's work that might allow them to verify or reject our conjectures.

\section{Beyond Bit-flip Efficiency}

Local Order Agnosticism is a general property of data structures and need not be limited to applications in bit-flip efficient methods. The techniques listed in this paper can be used for compression, error correction, or other coding theory applications. What we have found is a source of ``free" bits arising from either a limited valid set of local data arrangements or from the equivalency of sets of arrangements. We were interested in applications of those bits to bit flip efficiency but those bits can as easily be employed towards other goals. Especially because bit flip efficient coding is a relatively understudied area in Coding Theory, we recommend follow up work looking into Local Order Agnostic codes for compression and error correction.

\bibliography{Thesis} 
\bibliographystyle{ieeetr}

\end{multicols}
\end{document}